\documentclass{llncs}

\usepackage[utf8]{inputenc}
\usepackage{graphicx}
\usepackage{siunitx}
\usepackage{aicescover}

\graphicspath{{./figs/}}

\usepackage{color}
\definecolor{TODO}{RGB}{209,28,36}
\definecolor{TOREAD}{RGB}{180,180,74}

\definecolor{mandarine}{RGB}{255,128,0}
\definecolor{flore}{RGB}{69,171,71}
\definecolor{banane}{RGB}{180,180,74}
\definecolor{gray}{RGB}{144,146,142}

\usepackage{amsfonts}
\usepackage{subfig}

\usepackage{listings}
\usepackage{courier}
\usepackage{MnSymbol}
\newcommand{\AlgoCond}[2]{\makebox[4.0cm][l]{\color{gray}\ttfamily(if b in #1..#2)}}

\begin{document}
\mainmatter              % start of the contributions

\title{Streaming Data from HDD to GPUs for Sustained Peak Performance}

\author{Lucas Beyer \and Paolo Bientinesi}

\institute{RWTH Aachen University,\\
Aachen Institute for advanced study in Computational Engineering Science, Germany\\
\{beyer,pauldj\}@aices.rwth-aachen.de}

\aicescoverpage

\maketitle

%
% Modify the bibliography environment to call for the author-year
% system. This is done normally with the citeauthoryear option
% for a particular contribution.
\makeatletter
\renewenvironment{thebibliography}[1]
     {\section*{\refname}
      \small
      \list{}%
           {\settowidth\labelwidth{}%
            \leftmargin\parindent
            \itemindent=-\parindent
            \labelsep=\z@
            \if@openbib
              \advance\leftmargin\bibindent
              \itemindent -\bibindent
              \listparindent \itemindent
              \parsep \z@
            \fi
            \usecounter{enumiv}%
            \let\p@enumiv\@empty
            \renewcommand\theenumiv{}}%
      \if@openbib
        \renewcommand\newblock{\par}%
      \else
        \renewcommand\newblock{\hskip .11em \@plus.33em \@minus.07em}%
      \fi
      \sloppy\clubpenalty4000\widowpenalty4000%
      \sfcode`\.=\@m}
     {\def\@noitemerr
       {\@latex@warning{Empty `thebibliography' environment}}%
      \endlist}
      \def\@cite#1{#1}%
      \def\@lbibitem[#1]#2{\item[]\if@filesw
        {\def\protect##1{\string ##1\space}\immediate
      \write\@auxout{\string\bibcite{#2}{#1}}}\fi\ignorespaces}
\makeatother
\begin{abstract}
In the context of the genome-wide association studies (GWAS), one has to solve long sequences of generalized least-squares problems; such a task has two limiting factors: execution time --often in the range of days or weeks-- and data management --data sets in the order of Terabytes.
We present an algorithm that obviates both issues.
By pipelining the computation, and thanks to a sophisticated transfer strategy, we stream data from hard disk to main memory to GPUs and achieve sustained peak performance; with respect to a highly-optimized CPU implementation, our algorithm shows a speedup of 2.6x.
Moreover, the approach lends itself to multiple GPUs and attains almost perfect scalability.
When using 4 GPUs, we observe speedups of 9x over the aforementioned implementation, and 488x over a widespread biology library.


\end{abstract}
\section{GWAS, their Importance and Current Implementations}

In a nutshell, the goal of a genome-wide association study (GWAS) is to find an association between genetic variants and a specific trait such as a disease~[\cite{nhgri2012}].
Since there is a tremendous amount of such genetic variants, the computation involved in GWAS takes a long time, ranging from days to weeks and even months~[\cite{diego2012c}].
In this paper, we look at \emph{OOC-HP-GWAS}, currently the fastest algorithm available, and show how it is possible to speed it up by exploiting the computational power offered by modern graphics accelerators.

The solution of GWAS boils down to a sequence of generalized least squares (GLS) problems involving huge amounts of data, in the order of Terabytes.
The challenge lies in sustaining GPU's performance, avoiding idle time due to data transfers from hard disk (HDD) and main memory.
Our solution, \emph{cuGWAS}, combines three ideas: the computation is pipelined through GPU and CPU, the transfers are executed asynchronously, and the data is streamed from HDD to main memory to GPUs by means of a two-level buffering strategy.
Combined, these mechanisms allow cuGWAS to attain almost perfect scalability with respect to the number of GPUs;
when compared to OOC-HP-GWAS and another widespread GWAS library, our code is respectively 9 and 488 times faster.

In the first section of this paper, we introduce the reader to GWAS and the computations involved therein.
We then give an overview of OOC-HP-GWAS, upon which we build cuGWAS, whose key techniques we explain in Section~3 and which we time in Section~4.
We provide some closing remarks in Section~5.

\subsection{Biological Introduction to GWAS}
The segments of the DNA that contain information about protein synthesis are called \emph{genes}.
They encode so-called \emph{traits}, which are features of physical appearance of the organism --like eye or hair color-- as well as internal features of the organism --like blood type or resistances to diseases.
The hereditary information of a species consists of all the genes in the DNA, and is called \emph{genome};
this can be visualized as a book containing instructions for our body.
Following this analogy, the letters in this book are called \emph{nucleotides}, and determining their order is referred to as \emph{sequencing} the genome.
Even though the genome sequence of every individual is different, within one species most of it (99.9\% for humans) stays the same.
When a single nucleotide of the DNA differs between two individuals of the same species, this difference is called a single-nucleotide polymorphism (\emph{SNP}, pronounced ``snip'') and the two variants of the SNP are referred to as its \emph{alleles}.

Genome-wide association studies compare the DNA of two groups of individuals.
All the individuals in the \emph{case group} have a same trait, for example a specific disease, while all the individuals in the \emph{control group} do not have this trait.
The SNPs of the individuals in these groups are compared; if one variant of a SNP is more frequent in the case group than in the control group, it is said that the SNP is \emph{associated} with the trait (disease).
In contrast with other methods for linking traits to SNPs, such as inheritance studies or genetic association studies, GWAS consider the whole genome~[\cite{nhgri2012}].

\subsection{The Importance of GWAS}\label{sec:gwas}

\begin{figure}[ht!]
  \centering
  \includegraphics[width=1.0\textwidth]{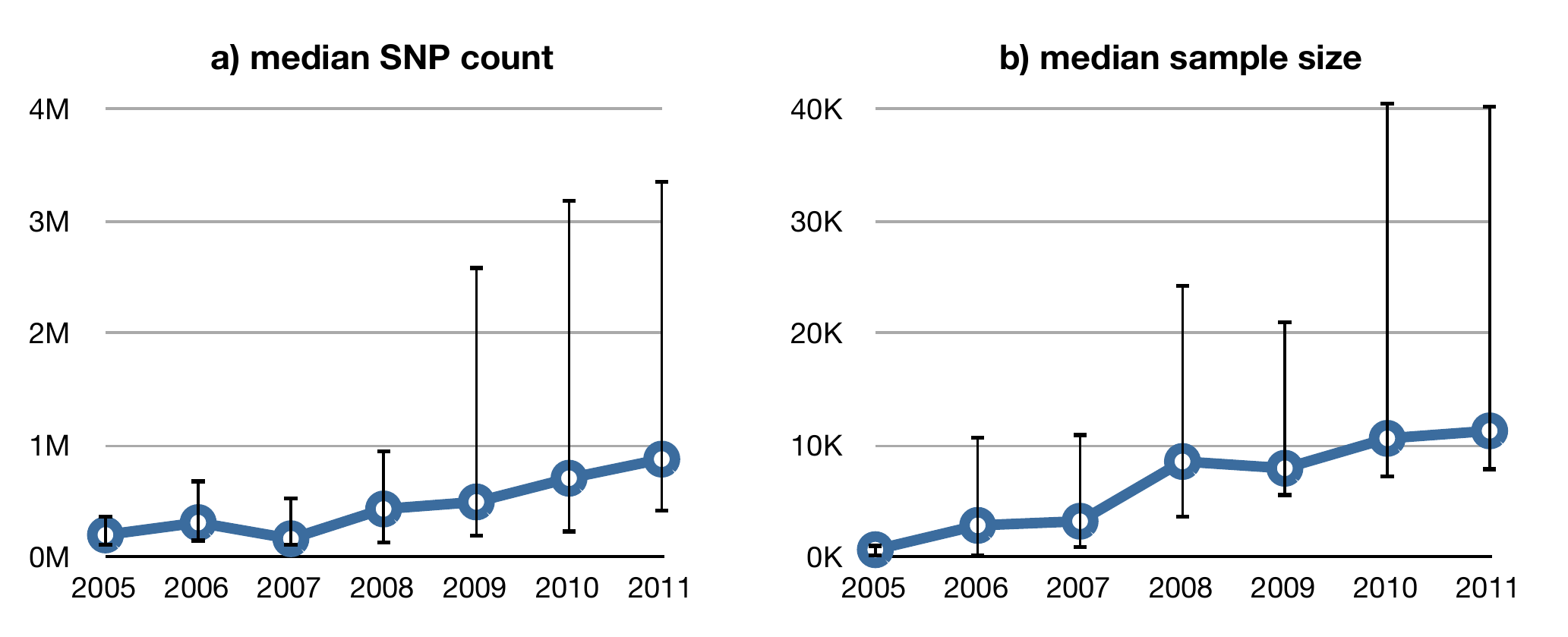}
  \caption[]{The median, first and second quartile of a) the SNP-count and b) the sample size of the studies each year.}
  \label{fig:gwas_sizes}
\end{figure}

We gathered insightful statistics about all published GWAS~[\cite{gwascatalog}].
Since the first GWAS started to appear in 2005 and 2006, the amount of yearly published studies has constantly increased, reaching more than 2300 studies in 2011.
This trend is summarized in the left panel of Fig.~\ref{fig:gwas_sizes}, showing the median SNP-count of each year's studies along with error-bars for the first and second quartiles.
One can observe that while GWA studies started out relatively small, since 2009 the amount of analyzed SNPs is growing tremendously.
Besides the number of SNPs, the other parameter relevant to the implementation of an algorithm is the \emph{sample size}, that is the total number of individuals of both the case and the control group.
What can be seen in Fig.~\ref{fig:gwas_sizes}b is that while it has grown at first, in the past four years the median sample size seems to have settled around 10\,000 individuals.
It is apparent that, in contrast to the SNP count, the growth of the sample size is negligible.
This data, as well as discussions with biologists, confirm the need for algorithms and software that can compute a GWAS with even more SNPs, and faster than currently possible.

\subsection{The Mathematics of GWAS}
The GWAS can be expressed as a variance component model~[\cite{diego2012b}] whose solution $r_i$ can be formulated as
\begin{equation}\label{eq:gls}r_i = (X_i^T M^{-1} X_i)^{-1} X_i^T M^{-1} y, \quad i=1..m\enspace,\end{equation}
where $m$ is in the millions and all variables on the right-hand side are known.
This sequence of equations is used to compute in $r_i$ the relations between variations in $y$ (the \emph{phenotype}\footnote{A phenotype is the observed value of a certain trait of an individual. For example, if the studied trait was the hair color, the phenotype of an individual would be the one of ``blonde'', ``brown'', ``black'' or ``red''.}) and variations in $X_i$ (the \emph{genotype}).
Each equation is responsible for one SNP, meaning that the number $m$ of equations corresponds to the number of SNPs considered in the study.

\begin{figure}[ht!]
  \centering
  \includegraphics[width=1.0\textwidth]{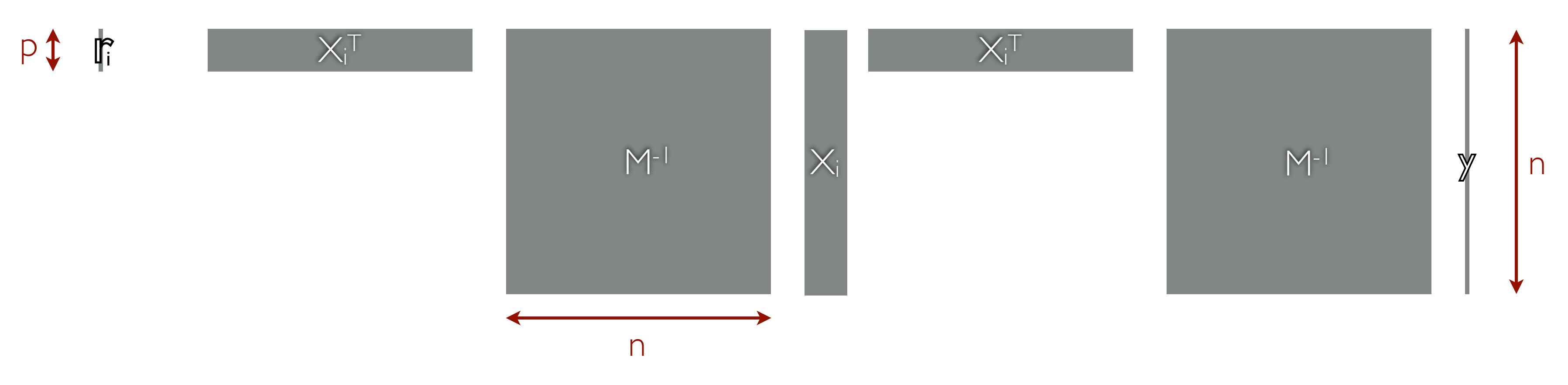}
  \caption[]{The dimensions of a single instance of (\ref{eq:gls}).}
  \label{dims_2d}
\end{figure}

Figure~\ref{dims_2d} captures the dimensions of the objects involved in one such equation.
The height $n$ of the matrices $X_i$ and $M$ and of the vector $y$ corresponds to the number of samples, thus each row in the \emph{design-matrix} $X_i \in \mathbb{R}^{n \times p}$ corresponds to a piece of each individual's genetic makeup (i.e. information about one SNP), and each entry in $y \in \mathbb{R}^n$ corresponds to an individual's phenotype.\footnote{In the example of the body height as a trait, the entries of $y$ would then be the heights of the individuals.}
$M \in \mathbb{R}^{n \times n}$ models the relations amongst the individuals, e.g. two individuals being in the same family.
Finally, an important feature of the matrices $X_i$ is that they can be partitioned as $(X_L | X_{R_i})$, where $X_L$ contains fixed covariates such as age and sex and thus stays the same for any $i$, while $X_{R_i}$ is a single column vector containing the genotypes of the $i$-th SNP of all considered individuals.

Even though (\ref{eq:gls}) has to be computed for every single SNP, only the right part of the design-matrix $X_{R_i}$ changes, while $X_L$, $M$ and $y$ stay the same.
\subsection{The Amount of Data and Computation Involved}\label{sec:related}
We analyze the storage size requirements for the data involved in GWAS.
Typical values for $p$ range between 4 and 20, but only one entry varies with $m$.
According to our analysis in Section~\ref{sec:gwas}, we consider $n=10\,000$ as the size of a study.
As of June 2012, the SNP database \emph{dbSNP} lists $187\,852\,828$ known SNPs for humans~[\cite{dbsnp}], so we consider $m=190\,000\,000$.
With these numbers, assuming that all data is stored as double precision floating point numbers.\footnote{Which may or may not be the optimal storage type. More discussion with biologists and analysis of the operations is necessary in order to find out whether {\tt float} is precise enough. If that was the case, the sizes should be halved.}
Therefore, the size of $y$ and $M$ is about 80\,MB and 800\,MB, respectively; both fit in main memory and in the GPU memory.
The output $r$ reaches 30\,GB, coming close to the limit of current high-end systems' main memory and is too big to fit in a GPU's 6\,GB of memory.
Weighting in at 14\,TB, $X$ is too big to fit into the memory of any system in the foreseeable future and has to be streamed from disk.

In the field of bioinformatics, the ProbABEL~[\cite{probabel2010}] library is frequently used for genome-wide association studies.
On a Sun Fire X4640 server with an Intel Xeon CPU 5160 (3.00 GHz), the authors report a runtime of almost 4~hours for a problem with $p=4$, $n=1500$ and $m=220\,833$, and estimate the runtime with $m=2\,500\,000$ to be roughly 43~hours\footnote{We only consider what the authors called the \emph{linear model} with the {\tt --mmscore} option as this solves the exact problem we tackle.} --almost two days.
Compared to the current demand, $m=2.5$~million is a reasonable amount of SNPs, but a population size of only $n=1500$ individuals is clearly much smaller than the present median (Fig.~\ref{fig:gwas_sizes}).
The authors state that the runtime grows more than linearly with $n$ and, in fact, tripling up the sample size from 500 to 1500 increased their runtime by a factor of 14.
Coupling this fact with the median sample size of about 10\,000 individuals, the computation time is bound to reach weeks or even months.
\section{Prior Work: the OOC-HP-GWAS Algorithm}
Presently, the fastest available algorithm for solving (\ref{eq:gls}) is OOC-HP-GWAS~[\cite{diego2012b}].
Since our work builds upon this CPU-only algorithm, we describe its salient features.

\subsection{Algorithmic Features}
OOC-HP-GWAS exploits the the symmetry and the positive definiteness of the matrix $M$, by decomposing it through a Cholesky factorization $L L^T = M$.
Since $M$ does not depend on $i$, this decomposition can be computed once as a preprocessing step and reused for every instance of (\ref{eq:gls}).
Substituting $L L^T = M$ into (\ref{eq:gls}) and rearranging, we obtain
\begin{equation}\label{eq:gls_chol}r_i = ( {\underbrace{(L^{-1} X_i)}_{\tilde{X}_i}}^T \underbrace{L^{-1} X_i}_{\tilde{X}_i})^{-1} {\underbrace{(L^{-1} X_i)}_{\tilde{X}_i}}^T \underbrace{L^{-1} y}_{\tilde{y}} \quad \text{for}~i=1..m\enspace,\end{equation}
effectively replacing the inversion and multiplication of $M$ with the solution of a triangular linear system ({\tt trsv}).

The second problem-specific piece of knowledge that is exploited by OOC-HP-GWAS is the structure of $X = (X_L | X_R)$: $X_L$ stays constant for any $i$, while $X_R$ varies; plugging $X_i = (X_L | X_{R_i})$ into (\ref{eq:gls_chol}) and moving the constant parts out of the loop leads to an algorithm that takes advantage of the structure of the sequence of GLS shown in Listing~\ref{lst:algo2}. The acronyms correspond to BLAS calls.

\begin{lstlisting}[label=lst:algo2,caption={Solution of the GWAS-specific sequence of GLS (\ref{eq:gls}).},mathescape,language=Python]
L   $\leftarrow$ potrf M						$(LL^T = M)$
Xl  $\leftarrow$ trsm L, Xl						$(\tilde{X}_L = L^{-1} X_L)$
y   $\leftarrow$ trsv L, y						$(\tilde{y} = L^{-1} y)$
rt  $\leftarrow$ gemv Xl, y						$(\tilde{r}_T = \tilde{X}_L^T \tilde{y})$
Stl $\leftarrow$ syrk Xl						$(S_{TL} = \tilde{X}_L^T \tilde{X}_L)$
for i in 1..m:
	Xri $\leftarrow$ trsv L, Xri				$(\tilde{X}_{R_i} = L^{-1} X_{R_i})$
	Sbl $\leftarrow$ dot Xri, Xl				$(S_{BL_i} = \tilde{X}_{R_i}^T \tilde{X}_L)$
	Sbr $\leftarrow$ syrk Xri					$(X_{BR_i} = \tilde{X}_{R_i}^T \tilde{X}_{R_i})$
	rb  $\leftarrow$ dot Xri, y					$(\tilde{r}_{B_i} = \tilde{X}_{R_i}^T \tilde{y})$
	r   $\leftarrow$ posv S, r					$(r_i = S_i^{-1} \tilde{r}_i)$
\end{lstlisting}

\subsection{Implementation Features}
Two implementation features allow OOC-HP-GWAS to attain near-perfect efficiency.
First, by packing multiple vectors $X_{R_i}$ into a matrix $X_{R_b}$, the slow BLAS-2 routine to solve a triangular linear system ({\tt trsv}) at Line 7 can be transformed into a fast BLAS-3 {\tt trsm}.
Then, Listing~\ref{lst:algo2} is an \emph{in-core} algorithm that cannot deal with an $X_R$ which does not fit into main memory.
This limitation is overcome by turning the algorithm into an \emph{out-of-core} one, in this case using a double-buffering technique:
While the CPU is busy computing the block $b$ of $X_R$ in a primary buffer, the next block $b+1$ can already be loaded into a secondary buffer through asynchronous I/O.
The full OOC-HP-GWAS algorithm is shown in Listing~\ref{lst:ooc}.
This algorithm attains more than 90\% efficiency.

\begin{lstlisting}[label=lst:ooc,caption=The full OOC-HP-GWAS algorithm.,language=Python,escapechar=!]
L   !$\leftarrow$! potrf M					!$(LL^T = M)$!
Xl  !$\leftarrow$! trsm L, Xl					!$(\tilde{X}_L = L^{-1} X_L)$!
y   !$\leftarrow$! trsv L, y					!$(\tilde{y} = L^{-1} y)$!
rt  !$\leftarrow$! gemv Xl, y					!$(\tilde{r}_T = \tilde{X}_L^T \tilde{y})$!
Stl !$\leftarrow$! syrk Xl					!$(S_{TL} = \tilde{X}_L^T \tilde{X}_L)$!
!\textcolor{banane}{\tt aio\_read}! Xr[1]
for b in 1..blockcount:
	!\textcolor{banane}{\tt aio\_read}! Xr[b+1]
	!\textcolor{banane}{\tt aio\_wait}! Xr[b]
	Xrb !$\leftarrow$! trsm L, Xrb			!$(\tilde{X}_b = L^{-1} X_b)$!
	for Xri in Xr[b]:
		Sbl !$\leftarrow$! gemm Xri, Xl		!$(S_{BL_i} = \tilde{X}_{R_i}^T \tilde{X}_L)$!
		Sbr !$\leftarrow$! syrk Xri			!$(X_{BR_i} = \tilde{X}_{R_i}^T \tilde{X}_{R_i})$!
		rb  !$\leftarrow$! gemv Xri, y		!$(\tilde{r}_{B_i} = \tilde{X}_{R_i}^T \tilde{y})$!
		r   !$\leftarrow$! posv S, r			!$(r_i = S_i^{-1} \tilde{r}_i)$!
	!\textcolor{banane}{\tt aio\_wait}! r[b-1]
	!\textcolor{banane}{\tt aio\_write}! r[b]
!\textcolor{banane}{\tt aio\_wait}! r[blockcount]
\end{lstlisting}

\section{Increasing Performance by Using GPUs}

While the efficiency of the OOC-HP-GWAS algorithm is satisfactory, the computations can be sped up even more by leveraging multiple GPUs.
With the help of a profiler, we determined (confirming the intuition), that the {\tt trsm} at line~10 in Listing~\ref{lst:ooc} is the bottleneck.
Since cuBLAS provides a high-performance implementation of BLAS-3 routines, {\tt trsm} it is the best candidate to be executed on GPUs.
In this section, we introduce cuGWAS, an algorithm for a single GPU, and then extend it to an arbitrary number of GPUs.

Before the {\tt trsm} can be executed on a GPU, the algorithm has to transfer the necessary data.
Since the size of $L$ is around 800~MB, the matrix can be sent once during the preprocessing step and kept on the GPU throughout the entire computation.
Unfortunately, the whole $X_R$ matrix weights in at several TB, way more than the 2~GB per buffer limit of a modern GPU.
The same holds true for the result $\tilde{X}_{R_b}$ of the {\tt trsm}, which needs to be sent back to main memory.
Thus, there is no other choice than to send it in a block-by-block fashion, each block ${X_{R_b}}$ weighting at most 2~GB.

\begin{figure}[ht!]
  \centering
  \includegraphics[width=0.8\textwidth]{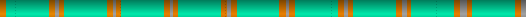}
  \caption[]{Profiled timings of the naïve implementation.}
  \label{fig:serial_gpu_timings}
\end{figure}

When profiled, a naïve implementation of the algorithm displays a pattern (Fig.~\ref{fig:serial_gpu_timings}) typical for applications in which GPU-offloading is an after-thought:
both GPU (\textcolor{flore}{green}) and CPU (\textcolor{gray}{gray}) need to wait for the data transfer (\textcolor{mandarine}{orange}); furthermore, the CPU is idle while the GPU is busy and vice-versa.

Our first objective is to make use of the CPU while the GPU computes the {\tt trsm}.
Regrettably, all operations following the {\tt trsm} (i.e. the for-loop at Lines~11--15 in Listing~\ref{lst:ooc}, which we will call the \emph{S-loop}) are dependent on its result and thus cannot be executed in parallel.
A way to break out of this dependency is to delay the S-loop by one block, in a pipeline fashion, so that the S-loop relative to the $b$-th block of $X_R$ is delayed and executed on the CPU, while the GPU executes the {\tt trsm} with the $(b\!+\!1)$-th block.
Thanks to this pipeline, we have broken the dependency and introduced more parallelism, completely removing the gray part of Fig.~\ref{fig:serial_gpu_timings}.

\subsection{Streaming Data from HDD to GPU}
The second problem with the aforementioned naïve implementation is the time wasted due to data transfers.
Modern GPUs are capable of \emph{overlapping} data transfers with computation.
If properly exploited, this feature allows us to eliminate any overhead, and thus attain sustained peak performance on the GPU.

The major obstacle is that the data is already being double-buffered from the hard-disk to the main memory.
A quick analysis shows that when targeting two layers of double-buffering (one layer for disk $\leftrightarrow$ main memory transfers and another layer for main memory $\leftrightarrow$ GPU transfers), two buffers on each layer are not sufficient anymore.
The idea here is to have two buffers on the GPU and three buffers on the CPU.

\begin{figure}
  \centering
  \includegraphics[width=0.8\textwidth]{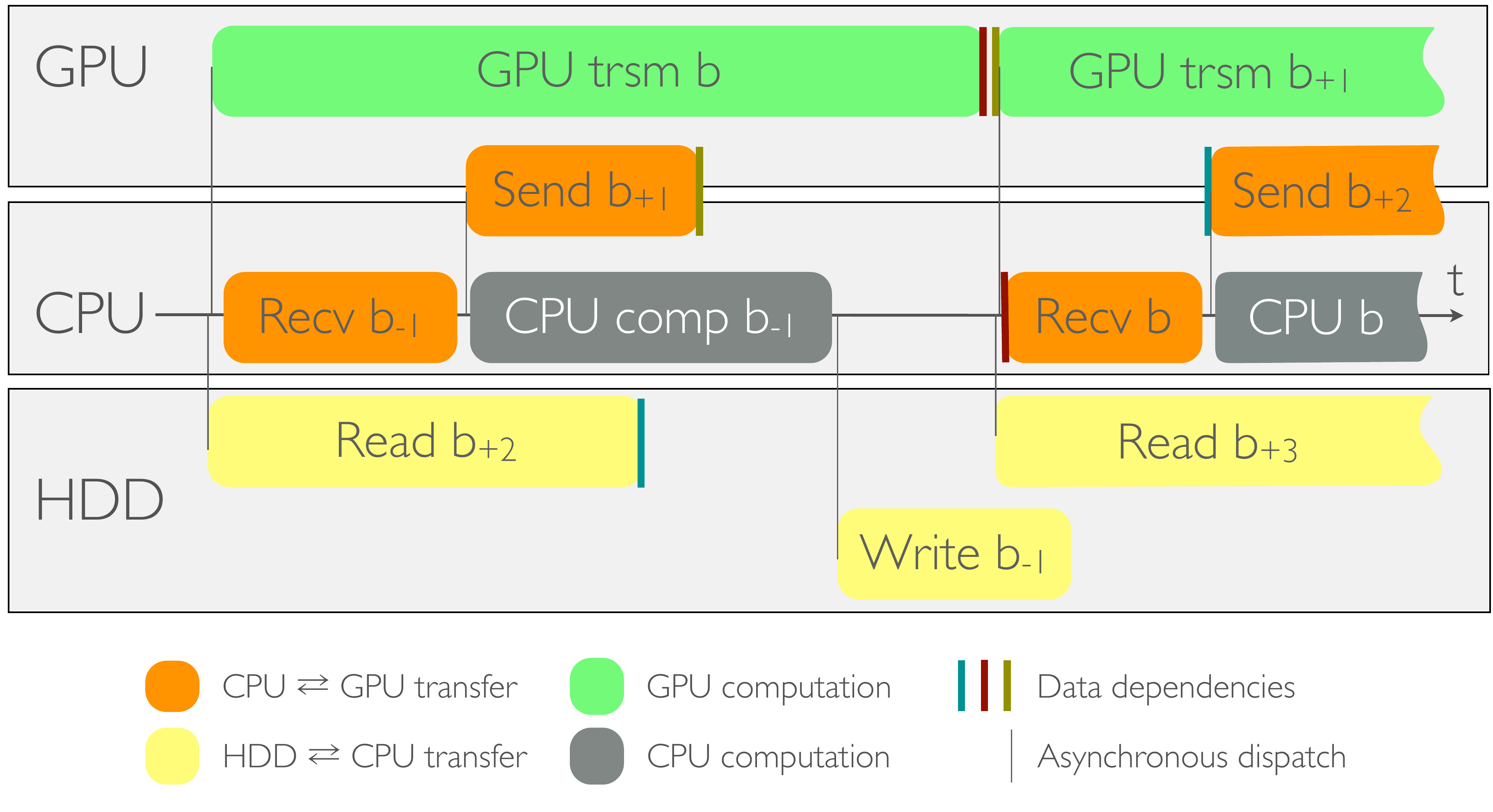}
  \caption[]{A task-perspective of the algorithm. Sizes are unrelated to runtime.}
  \label{fig:algo_timeline}
\end{figure}

The double-triple buffering can be illustrated from two perspectives: the tasks executed and the buffers involved. The former is presented in Fig.~\ref{fig:algo_timeline}; we refer the reader to~[\cite{diploma}] for a thorough description. Here we only discuss the technique in terms of buffers.

% TODO TODO
In this single-GPU scenario, the size of the blocks $X_{R_b}$ used in the GPU's computation is equal to that on the CPU.
When using multiple GPUs, this will not be the case anymore, as the CPU loads one large block and distributes portions of it to the GPUs.

The GPU's buffers are used in the same way as the CPU's buffers in the simple CPU-only algorithm:
While one buffer $\alpha$ is used for the computation, the data is transferred to and from the other buffer $\beta$.
But on the CPU's level (i.e. in RAM), three buffers are now necessary.
For the sake of simplicity, we avoid the explanation of the initial and final iterations and start with iteration $b$.

\begin{figure}[!ht]
  \centering
  \subfloat[Retrieve the previous result $b-1$ from GPU, and the second-next block $b+2$ of data from disk.]{
    \label{fig:buffers:a}
    \includegraphics[width=0.45\textwidth]{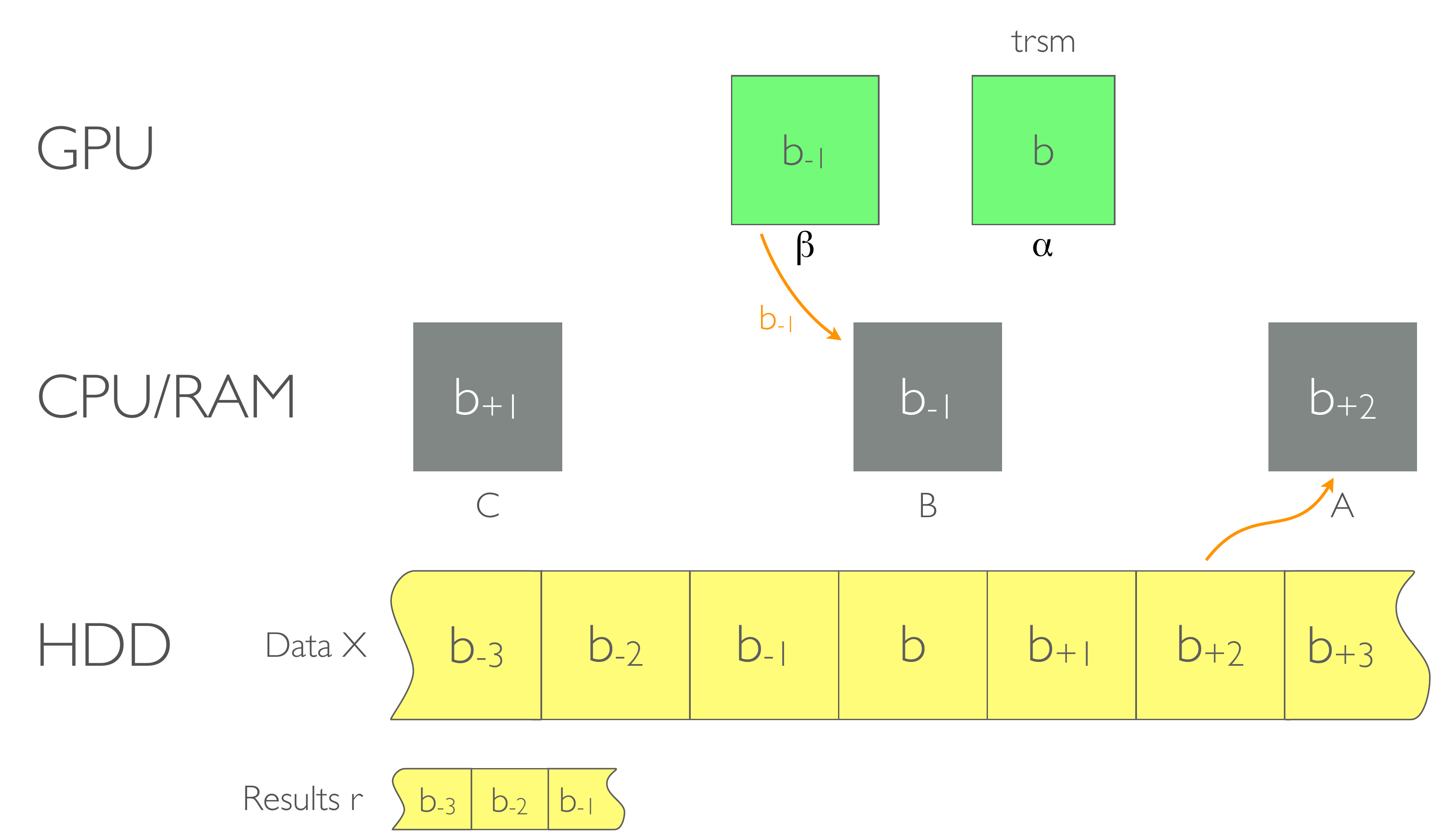}
  }
  \hfill
  \subfloat[Send the next block $b+1$ from RAM to the GPU, execute the S-loop on $b-1$ on the CPU.]{
    \label{fig:buffers:b}
    \includegraphics[width=0.45\textwidth]{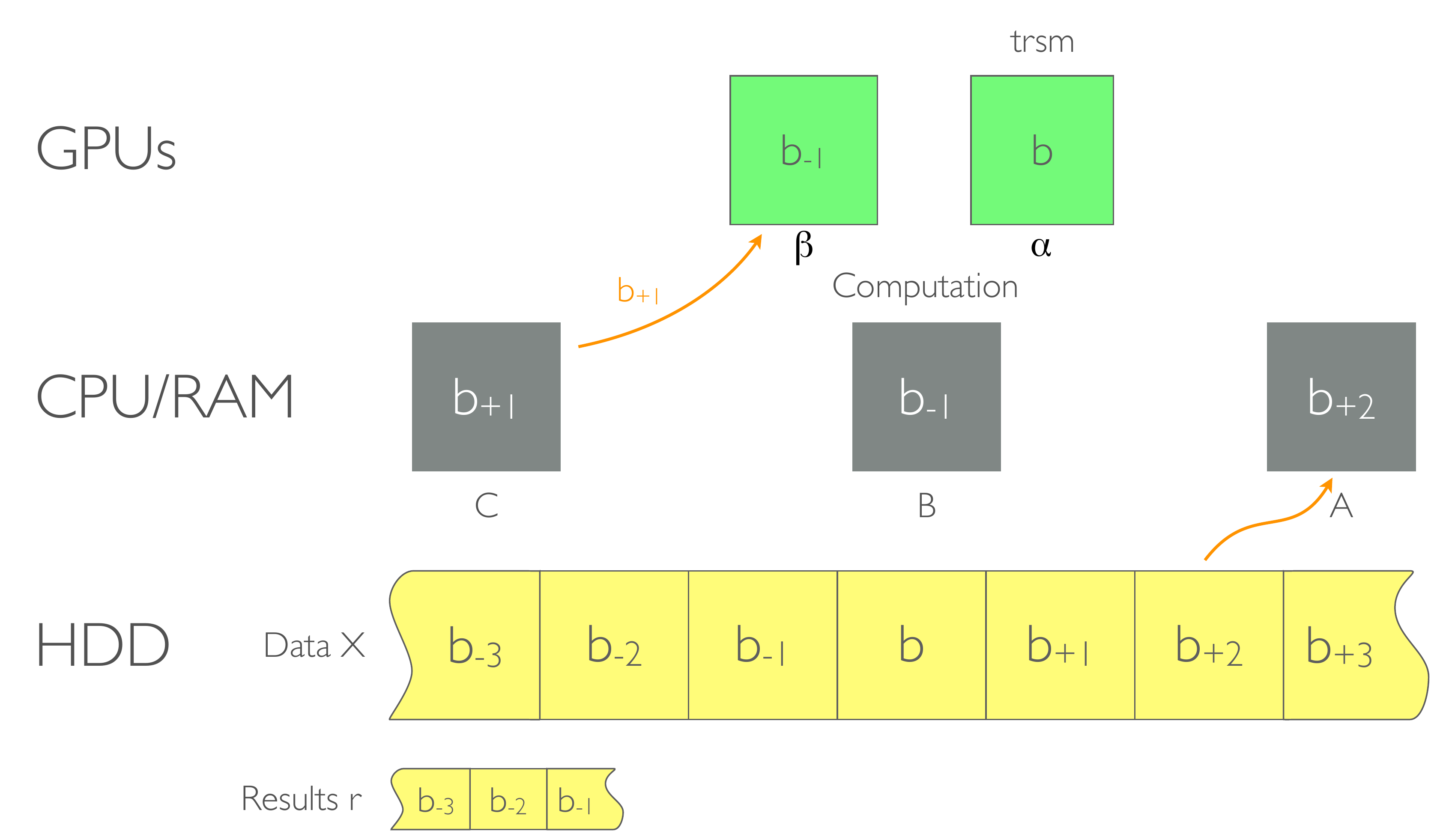}
  }
  \hfill
  \subfloat[Write the results $b-1$ to disk.]{
    \label{fig:buffers:c}
    \includegraphics[width=0.45\textwidth]{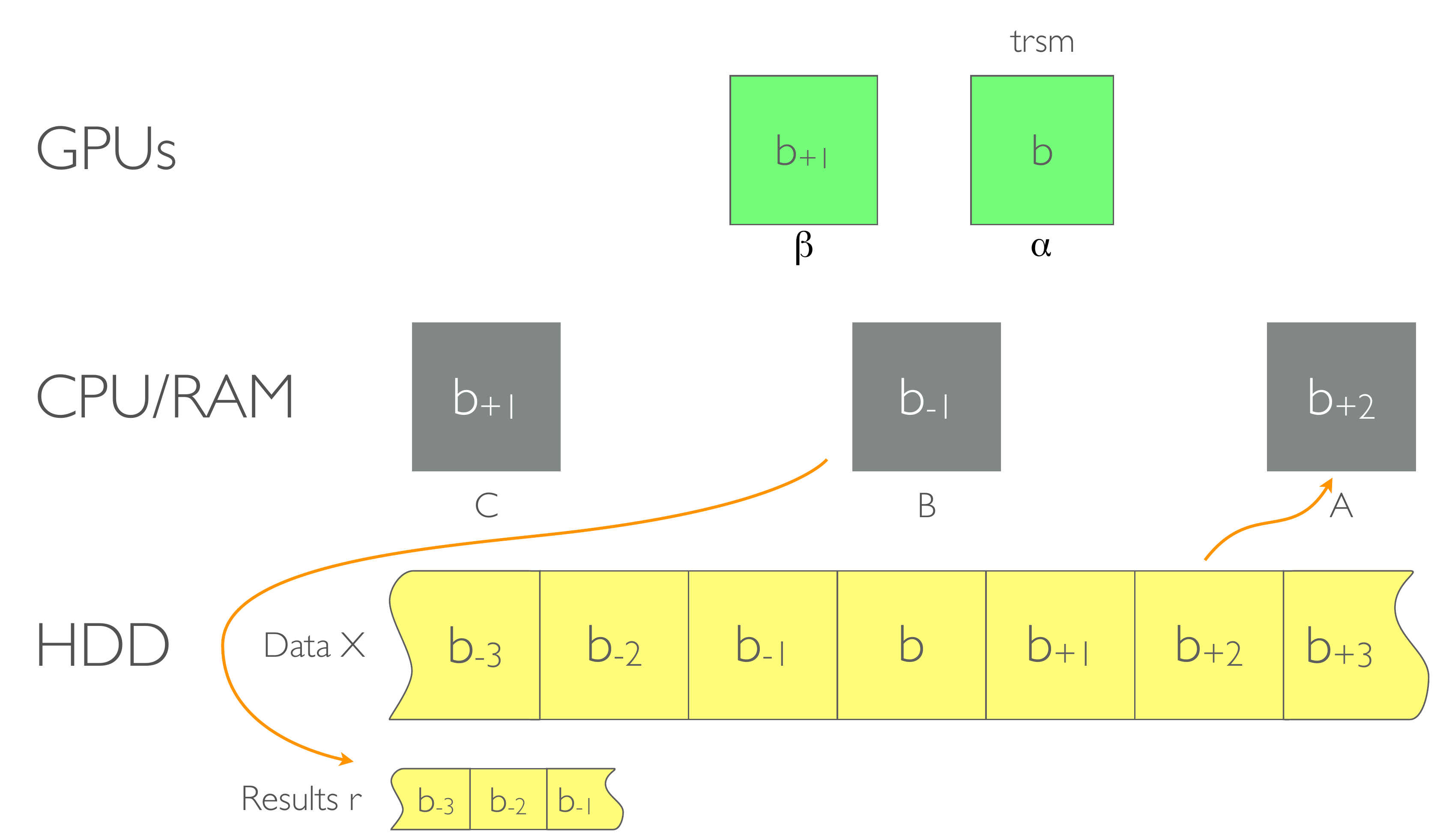}
  }
  \hfill
  \subfloat[Switch buffers at both levels for the next iteration.]{
    \label{fig:buffers:d}
    \includegraphics[width=0.45\textwidth]{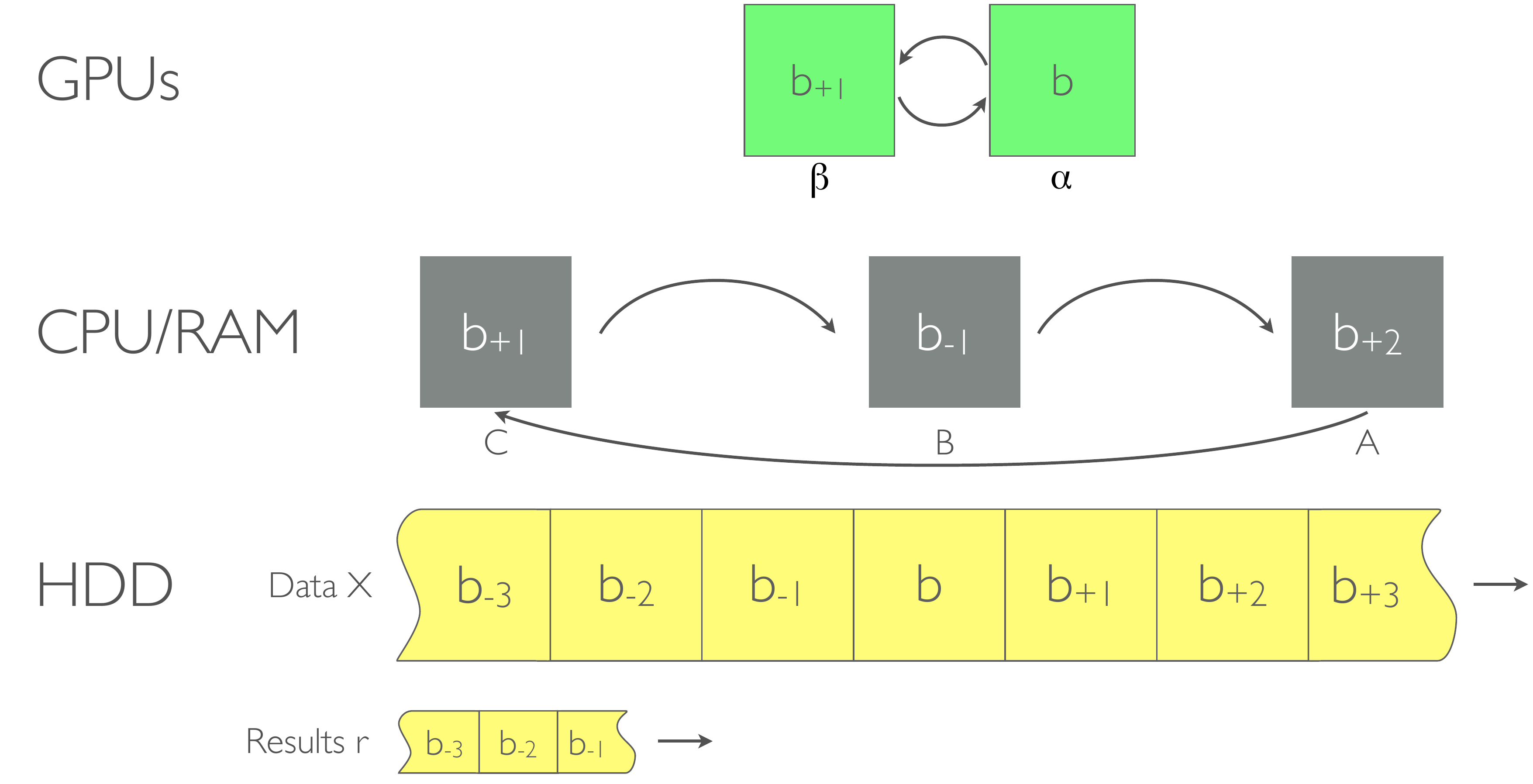}
  }
  \caption[]{The multi-buffering algorithm as seen from a buffer perspective.}
  \label{fig:buffers}
\end{figure}

With reference to Fig.~\ref{fig:buffers:a}, assume that the $(b\!-\!1)$-th, $b$-th and $(b\!+\!1)$-th blocks already reside in the GPU buffers $\beta$, $\alpha$, and in the CPU buffer $C$, respectively.
The block $b-1$ (i.e. buffer $\beta$) contains the solution of the {\tt trsm} of block $b-1$.
At this point, the algorithm proceeds by \emph{dispatching} both the read of the second-next block $b+2$ from disk into buffer $A$ and the computation of the {\tt trsm} on the GPU on buffer $\alpha$, and by receiving the result from buffer $\beta$ into buffer $B$.
The first two operations are \emph{dispatched}, i.e. they are executed asynchronously by the memory system and the GPU, while the last one is executed synchronously because these results are needed immediately in the following step.

As soon as the synchronous transfer $\beta \rightarrow B$ completes, the transfer of the next block $b+1$ from CPU buffer $C$ to GPU buffer $\beta$ is \emph{dispatched}, and the S-loop is executed on the CPU for the previous block $b-1$ in buffer $B$ on the CPU (see Fig.~\ref{fig:buffers:b}).

As soon as the CPU is done computing the S-loop, its results are written to disk (Fig.~\ref{fig:buffers:c}).
Finally, once all transfers are done, buffers are rotated (through pointer or index rotations, not copies) according to Fig.~\ref{fig:buffers:d}, and the loop continues with $b \leftarrow b+1$.

\subsection{Using Multiple GPUs}
This multi-buffering technique achieves sustained peak performance on one GPU.
Since boards with many GPUs are becoming more and more common in high-performance computing, we explain here how our algorithm is adapted to take advantage of all the available parallelism.
The idea is to increase the size of the $X_{R_b}$ blocks by a factor as big as the number of available GPUs, and then split the {\tt trsm} among these GPUs.
% TODO TODO
As long as solving a {\tt trsm} on the GPU takes longer than loading a large enough block $X_{R_b}$ from HDD to CPU, this parallelization strategy holds up to any number of GPUs.
Since in our systems loading the data from HDD was an order of magnitude faster than the computation of the {\tt trsm}, the algorithm scales up to more GPUs than were available. 
Listing~\ref{lst:multigpupar} shows the final version of cuGWAS.\footnote{The conditions for the first and last pair of iterations are provided in parentheses on the right.}

\section{Results}
In order to show the speedups obtained with a single GPU, we compare the hybrid CPU-GPU algorithm presented in Listing~\ref{lst:multigpupar} using one GPU with the CPU-only OOC-HP-GWAS.
Then, to determine the scalability of cuGWAS, we compare its runtimes when leveraging 1, 2, 3 and 4 GPUs.

In all of the timings, the time to initialize the GPU and the preprocessing (Lines~1--7 in Listing~\ref{lst:multigpupar}), both in the order of seconds, have not been measured.
The GPU usually takes 5\,s to fully initialize, and the preprocessing takes a few seconds too, but depends only on $n$ and $p$.
This omission is irrelevant for computations that run for hours.

\begin{lstlisting}[label=lst:multigpupar,caption={cuGWAS. The black bullet is a placeholder for ``all GPUs''.},language=Python,escapechar=!]
L   !$\leftarrow$! potrf M					!$(LL^T = M)$!
!\textcolor{mandarine}{\tt cublas\_send}! L !$\rightarrow$! L_gpu!$_\bullet$!
Xl  !$\leftarrow$! trsm L, Xl					!$(\tilde{X}_L = L^{-1} X_L)$!
y   !$\leftarrow$! trsv L, y					!$(\tilde{y} = L^{-1} y)$!
rt  !$\leftarrow$! gemv Xl, y					!$(\tilde{r}_T = \tilde{X}_L^T \tilde{y})$!
Stl !$\leftarrow$! syrk Xl					!$(S_{TL} = \tilde{X}_L^T \tilde{X}_L)$!
gpubs !$\leftarrow$! blocksize/ngpus
for b in -1..blockcount+1:
	!\textcolor{flore}{\tt cu\_trsm\_wait}! !$\alpha_{\bullet}$! !\hspace{2.51cm}! !\AlgoCond{1}{blockcount}!
    !\textcolor{mandarine}{\tt cu\_send\_wait}! C!$_\bullet$! !$\rightarrow \beta_{\bullet}$! !\hspace{1.67cm}! !\AlgoCond{2}{blockcount+1}!
	!$\alpha_\bullet \leftarrow$! !\textcolor{flore}{cu\_trsm\_async}! L_gpu!$_\bullet$!, !$\alpha_\bullet$!  !\AlgoCond{1}{blockcount}! !$(\tilde{X}_b = L^{-1} X_b)$!
    !\textcolor{banane}{\tt aio\_read}! Xr[b+2] !$\rightarrow$! A !\hspace{1.35cm}! !\AlgoCond{-1}{blockcount-2}!
    for gpu in 0..ngpus: !\hspace{1.35cm}! !\AlgoCond{2}{blockcount+1}!
	    !\textcolor{mandarine}{\tt cu\_recv}! B[gpu*gpubs..(gpu+1)*gpubs] !$\leftarrow \beta_{gpu}$!
	!\textcolor{banane}{\tt aio\_wait}! Xr[b+1] !$\rightarrow$! C !\hspace{1.35cm}! !\AlgoCond{0}{blockcount-1}!
    for gpu in 0..ngpus: !\hspace{1.35cm}! !\AlgoCond{0}{blockcount-1}!
    	!\textcolor{mandarine}{\tt cu\_send\_async}! C[gpu*gpubs..(gpu+1)*gpubs] !$\rightarrow \beta_{gpu}$!
	for Xri in B: !\hspace{2.7cm}! !\AlgoCond{2}{blockcount+1}!
		Sbl !$\leftarrow$! gemm Xri, Xl		!$(S_{BL_i} = \tilde{X}_{R_i}^T \tilde{X}_L)$!
		Sbr !$\leftarrow$! syrk Xri			!$(X_{BR_i} = \tilde{X}_{R_i}^T \tilde{X}_{R_i})$!
		rb  !$\leftarrow$! gemv Xri, y		!$(\tilde{r}_{B_i} = \tilde{X}_{R_i}^T \tilde{y})$!
		r   !$\leftarrow$! posv S, r			!$(r_i = S_i^{-1} \tilde{r}_i)$!
	!\textcolor{banane}{\tt aio\_wait}! r[b-2] !\hspace{2.4cm}! !\AlgoCond{1}{blockcount+1}!
	!\textcolor{banane}{\tt aio\_write}! r[b-1] !\hspace{2.21cm}! !\AlgoCond{1}{blockcount+1}!
    swap_buffers
\end{lstlisting}

\subsection{Single-GPU Results}
The experiments with a single-GPU were performed on the \emph{Quadro} cluster at the RWTH Aachen University; the cluster is equipped with two nVidia Quadro 6000 GPUs and two Intel Xeon X5650 CPUs per node.
The GPUs, which are powered by Fermi chips, have 6\,GB of RAM and a theoretical double-precision computational power of 515\,GFlops each.
In total, the cluster has a GPU peak of 1.03\,TFlops.
The CPUs, which have six cores each, amount to a total of 128\,GFlops and are supported by 24\,GB of RAM.
The cost of the combined GPUs is estimated to about \$10\,000 while the combined CPUs cost around \$2000.

Figure~\ref{fig:timings}a shows the runtime of OOC-HP-GWAS along with that of cuGWAS, using one GPU.
Thanks to our transfer-overlapping strategy, we can leverage the GPU's peak performance and achieve a 2.6x speedup over a highly-optimized CPU-only implementation.
cuBLAS' {\tt trsm} implementation attains about 60\,\% of the GPU's peak performance, i.e. about 309\,GFlops~[\cite{volkov}].
The peak performance of the CPU in this system amounts to 128\,GFlops; if the whole computation were performed on the GPU at {\tt trsm}'s rate, the largest speedup possible would be 2.4.
We achieve 2.6 because the computation is pipelined: the S-loop is executed on the CPU, in perfect overlap with the GPU. This means that the performance of cuGWAS is perfectly in line with the theoretical peak.

In addition, the figure indicates that the algorithm (1) has linear runtime in $m$ and (2) allows us to cope with an arbitrary $m$.
The red vertical line in the figure marks the largest value of $m$ for which two blocks of $X_R$ fit into the GPU memory for $n=10\,000$.
Without the presented multi-buffering technique, it would not be possible to compute GWAS with more than $m=22\,500$ SNPs, while cuGWAS allows the solution of GWAS with any given amount of SNPs.
\subsection{Scalability with Multiple GPUs}
To experiment with multiple GPUs, we used the \emph{Tesla} cluster at the Universitat Jaume I in Spain, since it is equipped with an nVidia Tesla S2050 which contains four Fermi chips (same model as the \emph{Quadro} system), for a combined GPU compute power of 2.06\,TFlops, but with only 3\,GB of RAM each.
The host CPU is an Intel Xeon E5440 delivering approximately 90\,GFlops.

\begin{figure}
  \centering
  \includegraphics[width=1.0\textwidth]{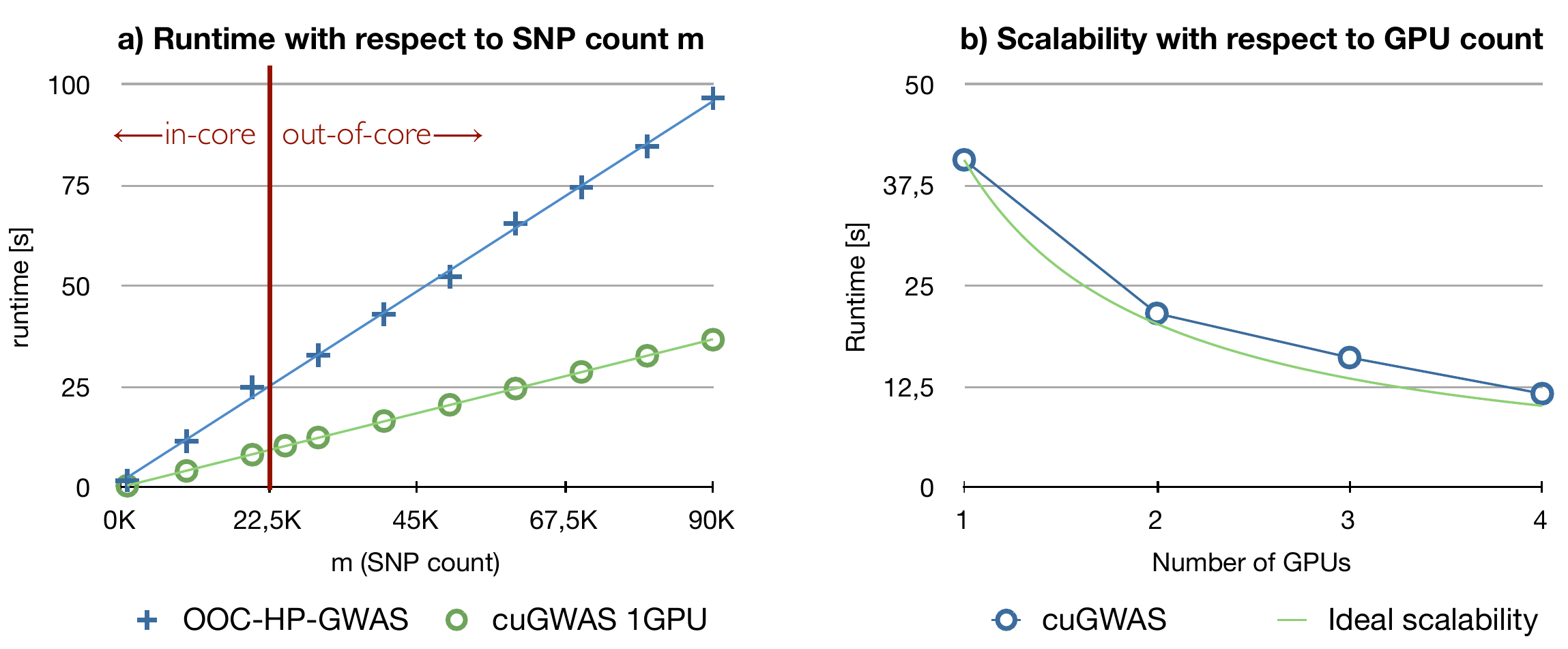}
  \caption{The runtime of our cuGWAS algorithm a) using 1GPU compared to OOC-HP-GWAS, using 1GPU and b) using a varying amount of GPUs.}
  \label{fig:timings}
\end{figure}

In order to evaluate the scalability of cuGWAS, we solved a GWAS with $p=4$, $n=10\,000$, and $m=100\,000$ on the \emph{Tesla} cluster, varying the number of GPUs.
As it can be seen in Fig.~\ref{fig:timings}b, the scalability of the algorithm with respect to the number of GPUs is almost ideal: Doubling the amount of GPUs reduces the runtime by a factor of $1.9$.

\section{Conclusion and Future Work}
We have presented a strategy which makes it possible to sustain peak performance on a GPU not only when the data is too big for the GPU's memory, but also for main memory.
In addition, we have shown how well this strategy lends itself to exploit an arbitrary number of GPUs.

As described by the developers of ProbABEL, the solution of a problem of the size described in Section~\ref{sec:related} by the GWFGLS algorithm took 4\,hours. In contrast, with cuGWAS we solved the same problem in 2.88\,s. Even accounting for about 6\,seconds for the initialization and Moore's Law (doubling the runtime as ProbABEL's timings are from 2010), the difference is dramatic. We believe that the contribution of cuGWAS is an important step towards making GWAS practical.

\subsubsection{Software}
The code implementing the strategy explained in this paper is freely available at \url{http://github.com/lucasb-eyer/cuGWAS} and \url{http://lucas-b.eyer.be}.

\subsubsection{Acknowledgements}
Financial support from the Deutsche Forschungsgemeinschaft (German Research Association) through grant GSC 111 is gratefully acknowledged.
The authors thank Diego Fabregat-Traver for providing us with the source-code of OOC-HP-GWAS, the Center for Computing and Communication at RWTH Aachen for the resources, Enrique S. Quintana-Ort\'{\i} for granting us access to the \emph{Tesla} system as well as Yurii S. Aulchenko for intorducing us to the computational challenges of GWAS.
%
% ---- Bibliography ----
%

\end{document}